\documentstyle[12pt]{article}
\advance\hoffset by -1.1truecm
\begin{document}
\title{Quantum Group Symmetric Bargmann Fock Construction\thanks{talk given
at the XXI-DGM Conference Tianjin, China in June 92} }
\author{Achim Kempf\thanks{supported by the German National Scholarship
Foundation}\\Sektion Physik der Ludwig-Maximilians-Universit{\"a}t\\
Theresienstr.37, 8000 M\"unchen 2
/ FRG\\{ \small E-mail:
 kempf@ls-wess.physik.uni-muenchen.dbp.de}}
\date{ }
\maketitle
\vskip-7.5cm
\hskip5.5cm
\sf Preprint LMU-TPW 92-25, \quad  hep-th/9211022 \rm
\vskip7.5cm

\begin{abstract}
Usually in quantum mechanics the Heisenberg algebra is generated by
operators of position and momentum. The algebra is then represented on
an Hilbert space of square integrable functions. Alternatively one generates
the Heisenberg algebra by the raising and lowering operators.
It is then natural to represent it on the Bargmann Fock space of holomorphic
functions. In the following I show that the Bargmann Fock construction
can also be done in the quantum group symmetric case. This leads
to a 'q- deformed quantum mechanics' in which the basic concepts, Hilbert
space of states and unitarity of time evolution, are preserved.
\end{abstract}

\section{Introduction}
There are already several approaches to $q$- deformed algebras of raising
and lowering operators in the literature [see e.g. 6,7,8,9,11].
Let us define a 'q- deformed' Heisenberg algebra generated by the operators
${a^{\dagger}}_i$ and $a^i$ by imposing the following commutation
relations which were
shown to be preserved under the action of the
quantum group $SU_q(n)$ [4,6]:
\begin{equation}
a^r a^s + c R_{ji}^{sr} a^j a^i = 0
\mbox{\quad,\quad}
a^r {a^{\dagger}}_s + 1/c R_{is}^{rj} {a^{\dagger}}_j a^i = {\delta}^r_s
\mbox{\quad,\quad}
{a^{\dagger}}_r {a^{\dagger}}_s + c R^{ji}_{sr} {a^{\dagger}}_j
{a^{\dagger}}_i = 0
\end{equation}
The R- matrix reads [for the foundations see e.g. 1,2]:
\begin{equation}
R = q \sum_i e^i_i \otimes e^i_i + \sum_{i\ne j} e^i_i \otimes e^j_j +
(q-1/q) \sum_{i>j} e^i_j \otimes e^j_i
\end{equation}
with $e^i_j \in M_n(C)$ matrix units.
In the bosonic case we have $c = -1/q$ and in the fermionic
case $c = q$, with $q$ a real number. Other commutation
relations are possible but the above relations are natural here in the sense
that the resulting algebra deviates the least possible from
the undeformed case [see 4]. Using the generalized twisting
method introduced in [10] it is easy to see, that no further continous
parameters can be introduced.

Since the ${a^{\dagger}}_i$ and $a^i$ will be represented as adjoint
operators we can recover hermitean position
and momentum operators:
\begin{equation}
x^i := (2 m \omega)^{-1/2} (a^{\dagger}_i + a^i) \mbox{ \quad , \quad }
p^i := i (2 m \omega)^{1/2} (a^{\dagger}_i - a^i)
\end{equation}
In the fermionic case one may consider a q- deformed system of $n$ spin $1/2$'s
in
a constant magnetic field in $z$- direction. With similar
linear combinations one then gets the $x$- and $y$- components
of the $i$'th spin.

We define the ground state as usual
$ < 0 \vert 0 > = 1 \mbox{\quad and\quad } {a^i \vert 0 >} = 0
\mbox{\quad for\quad } i = 1,...,n $.
The scalar product $< , >$ can then be shown to be still positiv
definit.

For the simplest Hamiltonian
$H := \omega N \mbox{ with } \omega \in {I\!\!R}^+$
, which is obviously scalar transforming and hermitean, the energy
spectrum is:
\begin{equation}
E_p = \omega (1 + c^{-2} + c^{-4} + ... + c^{-2(p-1)}) =
\omega \frac{c^{-2p} - 1}{c^{-2} - 1} \mbox{ \quad with \quad }
p = 0,1,2,...
\label{spec}
\end{equation}
\section{Bargmann Fock representation}
The operators  $a^i$ and $a^{\dagger}_i$ can be represented as
differentiation and
multiplication operators on a space of (deformed) holomorphic functions [4,5]:
$$ \rho : a^i \rightarrow {{\partial}_{\bar{\eta}}}^i
\mbox{ \qquad , \qquad } \rho : {a^{\dagger}}_i \rightarrow {\bar{\eta}}_i$$
Wave functions are the polynomials exclusively in ${\bar{\eta}}$'s
, which shall also be denoted holomorphic functions.
The bar operation is an antialgebra mapping i.e. we have for example
$\overline{{{\partial}_{\bar{\eta}}}^i {\bar{\eta}}_j} =
 {\eta}^j {{\partial}_{\eta}}_i $. The commutation relations among the
 ${\eta}^i, {\bar{\eta}}_i, {{\partial}_{\eta}}_i$ and
 ${{\partial}_{\bar{\eta}}}^i$ are not unique. I use a choice [4]
 that is
natural in the above mentioned sense, and which is similar but not
identical to the
differential calculus of Wess and Zumino [3].

The \it evaluation \rm of differentiation is defined as follows:
All ${{\partial}_{\bar{\eta}}}$'s are
to be commuted to the right, the ${{\partial}_{\eta}}$'s to the left. When
they arrive, the corresponding terms are to be set equal zero. What remains
is the value of the differentiation.

Let us define the '$c$- deformed' exponential function
\begin{equation}
{e_c}^{({{\partial}_{\eta}}_i {{\partial}_{\bar{\eta}}}^i)} := \sum\limits_{r}
\frac{{({{\partial}_{\eta}}_i {{\partial}_{\bar{\eta}}}^i)}^r}{[r]_c!}
\mbox{\qquad with \qquad}
[r]_c := \frac{c^{2r} - 1}{c^{2} - 1} .
\end{equation}
The scalar product $(,)$ of the wave functions $\phi$ and $\psi$ (which
are polynomials in $\bar{\eta}$) is then
given [4] by the
following "integral":
\begin{equation}
( \phi , \psi ) := {( \bar{\phi} {e_c}^{({{\partial}_{\eta}}_i
{{\partial}_{\bar{\eta}}}^i)} \psi )}_{\mbox{\small evaluated at \rm} \eta =
0 = \bar{\eta} }
\end{equation}
The evaluation procedure is as follows: At first the
differentiations are to be evaluated. Then all terms containing $\eta$'s
and ${\bar{\eta}}$'s
are to be set equal zero. Thus the result is a number. The new integral,
which can of course also be used in the undeformed theory, has
the same evaluation procedure for the bosonic as for the fermionic case.

It can be shown, that the operators $a^k$ and $a^{\dagger}_k$ are
adjoint in respect to $(,)$ i.e.
$({\rho}({a^{\dagger}}_k) \phi , \psi )   =  ( \phi , {\rho}(a^k) \psi )$
and that the wave function of the ground state, $1$, is normalized.
Thus the scalar product $(,)$ coincides with the bracket $ <,> $ of the
Fock space.
The Hilbert space of all wave functions is defined to be the set of power
series in ${\bar{\eta}}$'s that are square integrable in respect to $(,)$.
Let us introduce a more familiar notation for the scalar product:
\begin{equation}
\int d{\bar{\eta}} d{\eta}  \bar{\phi}
 e_c^{({\partial}_{{\eta}^i}{\partial}_{\bar{\eta}}^i)} \psi :=
{( \bar{\phi} {e_c}^{({{\partial}_{\eta}}_i {{\partial}_{\bar{\eta}}}^i)}
\psi )}_{\mbox{\small evaluated at } \eta = 0 = \bar{\eta} \rm}
\end{equation}

Like in the undeformed case it is now possible to represent every
operator $P$
on the Fock space also as an integral
kernel $G_P({{\bar{\eta}}}^{\prime},{\eta})$:
\begin{equation}
\int d{\bar{\eta}} d{\eta}  G_P({{\bar{\eta}}}^{\prime},{\eta})
 e_c^{({\partial}_{{\eta}^i}{\partial}_{\bar{\eta}}^i)}  {\psi}({\bar{\eta}})
= P{\psi}({\bar{\eta}}^{\prime})
\end{equation}
With natural commutation relations between different copies
of the function space (e.g. primed and unprimed)
the general rule for getting the integral kernel of an arbitrary
normal ordered operator $:P(a^{\dagger},a):$ turns out to be as follows:
Starting with the integral kernel of the identity operator
$e_c^{({{\bar{\eta}}^{\prime}}_i{\eta}^i)}$ one writes
for each $a^{\dagger}_i$ a ${{\bar{\eta}}^{\prime}}_i$ to the
left of $e_c^{({{\bar{\eta}}^{\prime}}_i{\eta}^i)}$ and for all
$a^i$ one writes ${\eta}^i$ to the right of the
$e_c^{({{\bar{\eta}}^{\prime}}_i{\eta}^i)}$.

The green function i.e. the integral kernel of the time
evolution operator $U = e^{-i(t_f-t_i)H}$  for the simple
Hamiltonian of section 1
is found to be
\begin{equation}
G_U =
 \sum_{r=0}^{\infty} \frac{ { ( {{\bar{\eta}}^{\prime}}_i{\eta}^i)}^r}{[r]_c!}
e^{-i\omega (t_f-t_i)[r]_{1/c}}
\end{equation}
\section{Introduction of driving forces}
Let us introduce a 'classical' but $q$-deformed driving
force $f(t)g^i$ where $f(t)$ denotes a complex- valued function describing
the time dependence of the driving force and $g$ shall be a constant unit
vector: ${\bar{g}}_ig^i = 1$. The Hamiltonian now reads:
\begin{equation}
H = \omega {\bar{\eta}}_i {\partial}^i_{\bar{\eta}} -
\bar{f}(t){\bar{g}}_i{\partial}^i_{\bar{\eta}} - f(t){\bar{\eta}}_ig^i
\end{equation}
Considering quantum mechanics as quantum field theory with zero space
dimensions, our driving forces are $q$-deformed Schwinger sources.
The algebra of the $g$'s and ${\bar{g}}$'s is now noncommutative, even
in the bosonic case [5].
Nevertheless, results that do only depend on the length of the force vector
are still ordinary
complex numbers for that the usual probability interpretation applies.
E.g. the vacuum-vacuum transition amplitude for the switching on
of a constant driving force can be calculated to be
\begin{eqnarray}
< 0 \vert 0(t_f,t_i)>  & =  & 1 + \frac{f\bar{f}}{{\omega}^2}
\sum_{z=2}^{\infty} \frac{(-i\omega)^z}{z!} (t_f - t_i)^z \nonumber\\
   & +  & \frac{(f\bar{f})^2}{{\omega}^4} \sum_{z=4}^{\infty}
\left\{ z-3 + \sum_{r=0}^{z-4} (z-3-r)([2]_{1/c})^{r+1} \right\}
\frac{(-i\omega)^z}{z!}
(t_f - t_i)^z \nonumber\\
  & +  & ...
\label{vacdef}
\end{eqnarray}
One recognizes that deviations from the undeformed case do not occur
before the second order in $(f \bar{f})$.
This is plausible because
the energy level of the first excitation is not deformed,
while the levels (see Eq.\ref{spec}) deviate the more, the
higher the excitation is.
If it is possible to extend this formalism to a quantum
field theory it can be expexted that the ultraviolet behaviour is strongly
influenced by the deformation parameter.
\smallskip\newline
\bf Acknowledgement: \rm

I would like to thank very much Prof. Jian-min Shen and Prof. Mo-Lin Ge for
the kind invitation to the conference and the generous hospitality.
\section{References}
\begin{enumerate}
\item L.D.Faddeev, N.Yu.Reshetikhin, L.A.Takhtajan,
Alg.Anal.1,1,(1989)178-206
\item S.Majid, Int.J.Mod.Phys. A. Vol. 5, No 1 (1990) 1-91
\item J.Wess, B. Zumino, Nucl. Phys. Proc. Suppl. 18B (1991) 302
\item A.Kempf, Preprint LMU-TPW 92-4, submitted to Lett. Math. Phys.
\item A.Kempf, Preprint LMU-TPW 92-7, submitted to J. Math. Phys.
\item W.Pusz, S.Woronowicz, Rep. Math. Phys. 27 (1989) 231.
\item L.Biedenharn, J. Phys. A 22 (1989) L 873
\item A.Macfarlane, J. Phys. A 22 (1989) 4581
\item D.B.Fairlie, C. Zachos, preprint ANL-HEP-CP-91-28 (1991)
\item A.Kempf, in Proc. XX DGM- Conference,
eds. S. Catto, A. Rocha (World Scientific, Singapore, 1991) p.546
\item M.Chaichian, P. Kulish, J. Lukierski, Phys. Lett. B 262 (1991) 43
\end{enumerate}
\end{document}